\documentclass[prb,aps,twocolumn]{revtex4}
\usepackage{graphicx}
\usepackage{amsmath}
\usepackage{amssymb}
\usepackage{bm}
\usepackage{ulem}

\newcommand{\beq}{\begin{equation}}
\newcommand{\eeq}{\end{equation}}
\newcommand{\bea}{\begin{eqnarray}}
\newcommand{\eea}{\end{eqnarray}}
\newcommand{\bei}{\begin{itemize}}
\newcommand{\eei}{\end{itemize}}

\newcommand{\ra}{\rangle}

\begin{document}

\title{Numerical Analysis of Quasiholes of the Moore-Read Wavefunction}

\author{M. Baraban$^1$, G. Zikos$^2$, N. Bonesteel$^2$, and S. H. Simon$^3$}
\affiliation{$^1$ Department of Physics, Yale University, 217 Prospect Street, New Haven, Connecticut 06511 \\
$^2$ Department of Physics and National High Magnetic Field
Laboratory, Florida State University, Tallahassee, FL 32310 \\
$^3$ Rudolf Peierls Centre for Theoretical Physics, Oxford University, 1 Keble Road, Oxford OX1 3NP
}

\begin{abstract}
We demonstrate numerically that non-Abelian quasihole (qh) excitations of
the $\nu = 5/2$ fractional quantum Hall state have some of the key
properties necessary to support quantum computation.  We find that as
the qh spacing is increased, the unitary transformation which
describes winding two qh's around each other converges
exponentially to its asymptotic limit and that the two orthogonal
wavefunctions describing a system with four qh's become
exponentially degenerate.  We calculate the length scales for these
two decays to be $\xi_{U} \approx 2.7 \, \ell_0$ and $\xi_{E} \approx
2.3 \, \ell_0$ respectively.  Additionally we determine which fusion
channel is lower in energy when two qh's are brought close
together.
\end{abstract}
\date{\today}
\pacs{
73.43.-f 
dimensions (anyons, composite fermions, Luttinger liquid, etc.) (for
anyon mechanism in superconductors, see 74.20.Mn) } \maketitle

The proposal to use quantum Hall states as a platform for quantum
computation has spurred a great deal of interest\cite{RevModPhys,Sarma,generalpapers}.
These quantum Hall systems are believed to have natural ``topological"
immunity to decoherence and therefore hold particular promise for
quantum computation.  In so-called non-Abelian quantum Hall systems,
the ground state is highly degenerate in the presence of
quasiparticles (qp's), and this degenerate space can be used to store quantum
information.  Operations on this space are then performed by
adiabatically dragging qp's around each other, thus
``braiding" their world-lines in 2+1 dimensions.

Although there is currently no definitive experimental evidence that
non-Abelian quantum Hall states even exist, the community now strongly
suspects\cite{RevModPhys} that the quantum Hall plateau observed at
Landau level (LL) filling fraction $\nu=5/2$ is the non-Abelian Moore-Read (MR)
phase\cite{MooreRead91} (or its closely related particle-hole
conjugate\cite{AntiPfaffian}).  While the MR phase is,
strictly speaking, not capable of universal topological quantum
computation (computation by braiding qp's around each other
at large distances), a scheme has been devised\cite{Bravyi} that in
principle allows error free quantum computation by supplementing these
topological processes with nontopological processes where
qp's are moved together and allowed to interact.
Furthermore, the MR phase is frequently viewed as the simplest
paradigm of a non-Abelian state of matter, and is therefore a logical
starting point for detailed analysis\cite{RevModPhys}.

In order for topological (or partially topological) schemes for
quantum computation to be scalable (i.e., to allow large scale quantum
computation), a number of crucial conditions must
hold\cite{RevModPhys}.  {\sl Condition} (1) {\sl As all of the
  qp's are moved apart from one another, the splitting of
  the energy levels of the putatively degenerate ground state space
  must converge to zero at least as fast as $e^{-R/\xi_E}$ where $R$
  is the minimum distance between qp's.}  In the literature,
there has been numerical work suggesting that condition (1) may not be
true\cite{Jain} for the MR state.  One of the goals of our
work is to perform more precise numerical calculations to determine
whether this numerical conclusion holds up to more careful scrutiny.
{\sl Condition} (2) {\sl As qp's are moved apart from each
  other, the unitary transformation that results from adiabatically
  dragging one qp around another must converge to its
  asymptotic limit at least as fast as $e^{-R/\xi_U}$}.  For the MR state, several theoretical arguments suggest that this is true \cite{Nayak,GurarieNayak,ReadRecent}; however, in these theoretical works the precise length scale $\xi_U$ remains unknown.  Presumably $\xi_E$ and
$\xi_U$ are both on the scale of a magnetic length multiplied by some
number of order unity.  If this number of ``order unity" happens to be
very large, it could in principle start to cause trouble for practical
implementation of topological schemes.  We will explicitly determine
both $\xi_U$ and $\xi_E$ numerically.  Finally, {\sl Condition} (3)
{\sl One must be able to measure the topological quantum number
  associated with a group of qp's}.  Proposals have been
made that such quantum numbers can be measured using
interferometry\cite{Sarma,Overbosch,RevModPhys}.  However, this scheme has
turned out to be very difficult experimentally.  Another possible way
to measure the topological quantum number of, say, two qp's,
is to move the qp's microscopically close and precisely
measure the force between them (or equivalently the energy change of
moving them).  While this may not sound any easier, it nonetheless
proposes a different route to making this measurement should
interferometry prove to be impossible.  In the current paper we will
attempt to numerically evaluate this energy change and show how it
reflects the quantum number of a pair of qp's.  See Ref.~\onlinecite{Pachos} for a similar
analysis of the Kitaev model.

\begin{figure}[ht!]
  \begin{center}
  \includegraphics[width=0.95\columnwidth]{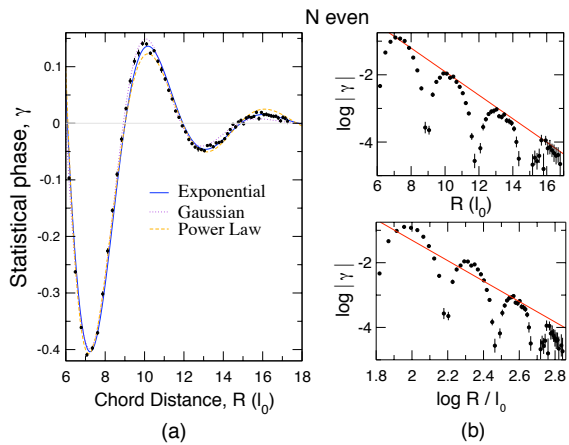}
  \end{center}
  \caption{ \label{fig:berryeven} (color online) Statistical phase for
    winding one qh around another in the spherical geometry
    as defined in the text. Data is shown for the case of even total
    numbers of electrons for which the topological quantum number of
    the pair of qh's is $1$.
    In (a) we have plotted the statistical phase versus the chord distance
    $R$ between the two qh's and fit the
    data using $\cos(a \frac{R}{l_0} + b)$ for the oscillatory part
    and a decay term that is either exponential, Gaussian or power law. We fit the data starting at $R =
    6.48 \,l_0$ and find the value of the reduced $\chi^2$ is smallest
    for exponential decay with a value of 1.42 while for power law and
    Gaussian decays it is 7.22 and 5.52, respectively. The good fit to
    exponential decay is also confirmed when we plot the absolute
    value of the data on log and log-log scales (b) and perform linear
    fits to the extrema of the oscillations. The linear fit is clearly
    much better on the log plot demonstrating that the oscillations
    decay exponentially rather than as a power law.}
\end{figure}

Our numerical work is performed on a spherical geometry with a
monopole of flux $N_{\phi}$ at the center of the sphere and $N$
electrons on the surface.  For the MR state\cite{MooreRead91},
$N_\phi$ is given by $N_{\phi} = 2 N -3 + n_{qh}/2$ and $n_{qh}$ is
the number of quasiholes (qp's with positive charge).  The
radius of the sphere is $(N_{\phi}/2)^{1/2} \ell_0$ where $\ell_0$ is
the magnetic length.  For the the purpose of stating the decay lengths
$\xi_{U}$ and $\xi_{E}$, the distance R between quasiholes (qh's) will be
written in terms of the chord length. The definitions of $\xi_E$ and $\xi_U$ are given below.  It should be noted that while their precise values depend on the the particular qh configurations used in our calculations, alternate definitions for different qh configurations will give results that only differ by factors of order unity.

We consider the MR wavefunction (wf) in the presence of qh's
which is defined as the zero energy space of a special short-ranged
three-body interaction\cite{Greiter}.  Although this is just a model
interaction, the ground state wf turns out to be an accurate
approximation for more realistic interactions\cite{PfaffianOverlaps}.  Thus our calculations are variational in nature.
Pairs of qh excitations carry the topological quantum number
``1" or ``$\psi$"
which represent the two states of a qubit, and the
degeneracy\cite{endnote1} of a system with $n_{qh}$ qh's is
$2^{\frac{n_{qh}}{2} - 1}$.


We start by considering the case of two qh's for which the
ground state is unique.  In this case we can address condition (2)
above by calculating the braiding statistics of these two
qh's. To do so in the spherical geometry, we compute the
Berry phase accumulated when one qh is moved
adiabatically around the equator of the sphere while the second
qh is held fixed first on the north pole and then on the
south pole. Both these Berry phases have contributions from the
statistical phase associated with the two qh's, and the
Aharanov-Bohm phase due to the applied magnetic field. To isolate
the statistical phase we therefore compute the {\it difference}
between these two phases. In the planar geometry this difference
would correspond to the change in the Berry phase when one qh
is moved in a closed loop while a second qh is held fixed
first inside the loop and then outside the loop.

The Berry phases are all calculated numerically using a
Monte Carlo method essentially identical to that described in
Ref.~\onlinecite{Tserkovnyak}. We use the MR wf with
two qh's, which is
not an exact wf for the realistic Coulomb interaction, but
is quite accurate nonetheless as prior numerical work has demonstrated\cite{PfaffianOverlaps}.   When we drag qh's we can think
of having added a highly localized potential well to the system whose
position moves as a function of time. However, since our Berry phase
calculation does not involve a detailed Hamiltonian per-se, our
results are independent of the form of this potential well.
(Further details of the methods used will
be given in Ref.~\onlinecite{Zikos}). For the cases of either an
even or odd number of electrons on the sphere, the two qh's
together must have topological quantum numbers 1 or $\psi$,
respectively.  The statistical phase is then
expected\cite{MooreRead91,ReadRecent,Nayak,RevModPhys} to converge
either to zero (if the quantum number is 1) or $\pi$ (if the
quantum number is $\psi$) as the distance between the qh's
is increased. Indeed, for an even number of electrons, we show in Fig.~\ref{fig:berryeven} that as the sphere
is made larger, the convergence is exponential and the
decay scale is roughly $\xi_U \approx 2.7 \, \ell_0$.  Similar results were obtained for the case of an odd number of electrons where the phase converges exponentially to $\pi$ with roughly the same decay scale.

The difference between the even and odd case can be interpreted as the
non-Abelian component -- i.e, the part of the phase that depends on
which topological sector the two qp's are in.  We conclude
that this non-Abelian contribution does indeed converge exponentially
with increasing system size as desired by condition (2).  (Ideally we would
like to determine the unitary transformation that occurs on this two
dimensional ground state space when particles are braided around each
other as in Ref.~\onlinecite{Tserkovnyak}.  However, we have found
that it is currently numerically too demanding to demonstrate
exponential convergence in this more complicated situation).

The oscillations in Fig.~\ref{fig:berryeven}
(and in the later Figures) are
not unexpected.  In the closely related system of a p-wave paired
superfluid, the oscillating form of the wavefunctions can be
calculated explicitly\cite{Gurarie,Cheng}.  However, in this quantum
Hall system those results would only be qualitative.

To address condition (1) above, we now turn to the case of four
qh's and restrict ourselves to an even number of
electrons. We implement a trial wf approach using the Moore-Read wf with qh's, which is the ground
state of a special three body interaction, but we will evaluate its
energy with a 1st excited LL Coulomb interaction\cite{Park}, for which the MR
wf is not the exact ground state. We nonetheless expect
this hybrid approach to give accurate results because the MR
wf is an extremely accurate approximation
of the exact ground state of the Coulomb interaction. Our calculation is also an exact statement about the lowest order
perturbation of the special three body interaction towards the
Coulomb point\cite{PfaffianOverlaps}.

\begin{figure}[htb!]
  \begin{center}
    \includegraphics[width=0.95\columnwidth]{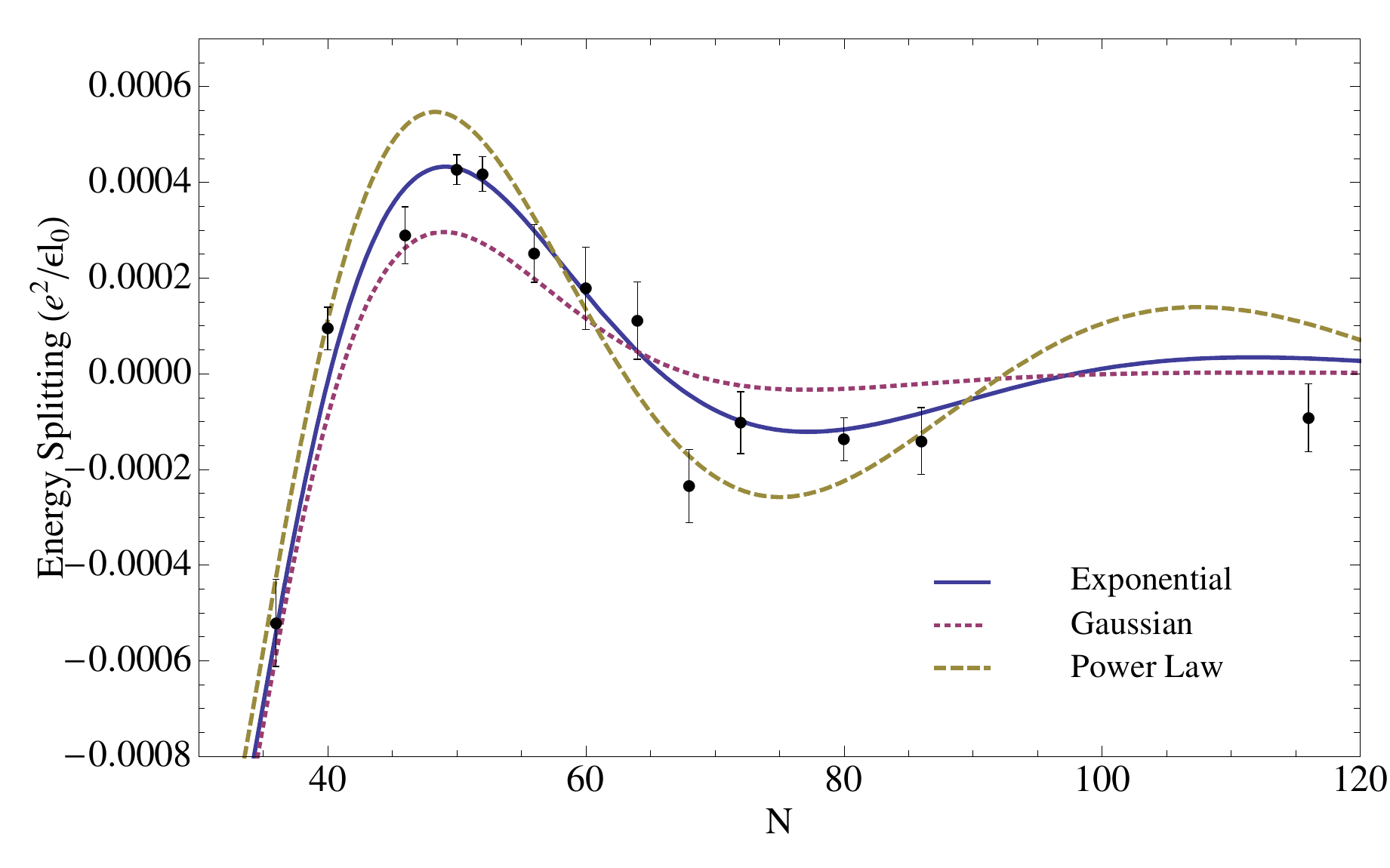}
  \end{center}
  \caption{ \label{fig:degeneratenergies} (color online) Energy
    splitting of the eigenstates on a sphere with four qh's as a
    function of system size.  The four qh's are placed on the
    corners of an equilateral tetrahedron.  The data is fit with the
    function $\cos(a \sqrt{N} + b)$ multiplied by an exponential, Gaussian, or power law decay function.
    The distance between particles grows as $\sqrt{N}$, so this is
    essentially the same fit as used in Fig.~\ref{fig:berryeven}.  The
    data is fit starting with $N=12$ (not shown because the fits are
    nearly identical at the small $N$ values).  The reduced $\chi^{2}$
    values for the exponential, Gaussian, and power law fits are 3.39,
    7.73, and 6.49 respectively, which helps confirm our expectation
    that two energies become exponentially degenerate as the
    qh's move apart.  We have not shown the log and log-log
    plots here because such plots discard the sign, and in the absence
    of a higher density of points, are hard to interpret.}
\end{figure}

For the MR wf with four qh's, there are two putatively degenerate ground
state wavefunctions\cite{MooreRead91,Nayak}.  Using a spherical geometry, we place four
qh's on the corners of an equilateral tetrahedron and implement
standard Monte Carlo procedures to evaluate the energy splitting
between the two eigenstates of the interaction within the two
dimensional ground state subspace (Details will be presented in
Ref.~\onlinecite{Zikos}).  Results are presented in Fig.~\ref{fig:degeneratenergies} as a
function of system size, and indeed it appears that the two blocks
become degenerate exponentially as the distance between qp's
increases as required by condition (2) with a decay length of $\xi_E
\approx 2.3 \, \ell_0$.
This result appears to contradict results of Ref.~\onlinecite{Jain} which claimed an
algebraic rather than exponential decay.  Although the methods used by
the two works are essentially identical, in Ref.~\onlinecite{Jain} the
MR state is studied in the lowest LL (where MR is known not
to be a good trial wf) whereas we have studied it in the first LL
where it is known to be a very good trial wf and thought to be experimentally relevant.
Differences with Ref.~\onlinecite{Jain} could occur because of differing
levels of Monte Carlo error as well.  The numerical
difficulty of collecting data is substantial, so admittedly our error
bars are currently somewhat larger than desirable.  However, we will
continue to collect data and these results will almost certainly
improve.

\begin{figure}[bt!]
  \begin{center}
    \includegraphics[width=0.95\columnwidth]{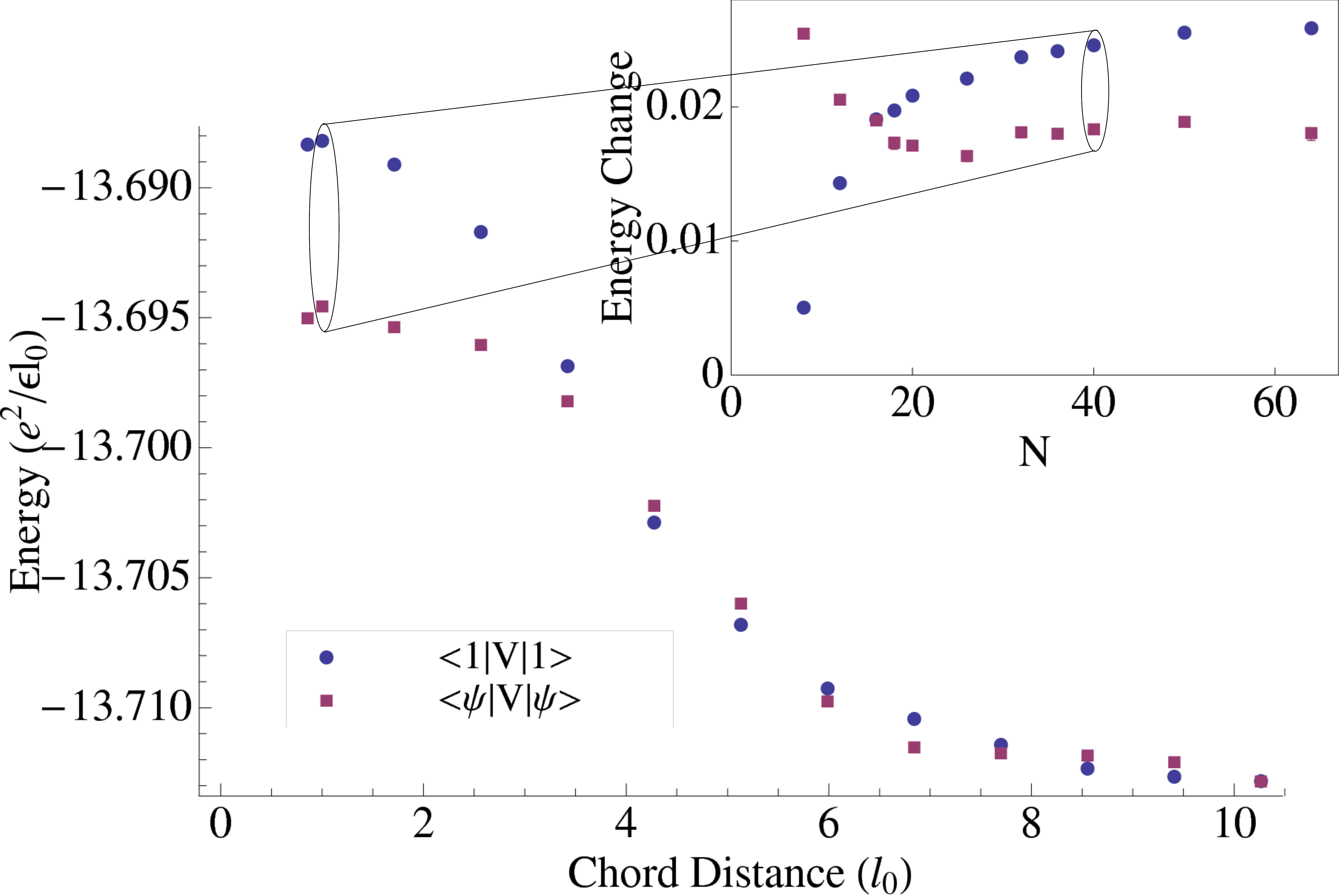}
  \end{center}
  \caption{ \label{fig:movingqhstogether} (color online) This figure
    shows the total energy of the $N=40$ system with four qh's
    as qh $2$ is moved closer to its pair, qh $1$.
    Qh $1$ is located at the north pole and qh $2$ is
    moved along a path that keeps a certain analytic form\cite{Nayak}
    for the trial states $|1 \ra$ and $|\psi \ra$ precisely
    orthogonal\cite{Zikos}.  The two states of the system, $|1\ra$ and
    $|\psi\ra$, which are nearly degenerate when the qh's are
    well separated, split as the qh's approach each other.  The
    inset shows the energy needed to move qh $2$ within
    $\ell_0$ of qh $1$ for different values of $N$.  We find
    that it takes more energy to bring the two qh's together
    when the system is in state $|1\ra$ than in state $|\psi \ra$, and
    that the energy splitting is on the order of $0.01 e^2/\epsilon
    \ell_0$.}
\end{figure}

Finally we turn to the issue of measurement, condition (3).  Here, we start with four qh's at the corners of a tetrahedron on a relatively large sphere ($N=40$) where the two ground state wavefunctions are close to degenerate.    We then move qh $2$ close to $1$ and observe the change in energy of the two wavefunctions.   It turns out (and we will show in detail in Ref.~\onlinecite{Zikos}) that if we choose to move the qh's together along an appropriately chosen path, then the conformal block wavefunctions defined in Ref. \onlinecite{Nayak} diagonalize the interaction.   These two conformal block wavefunctions, known as $|1\ra$ and $|\psi\ra$, are constructed such that the pair of qh's $1$ and $2$ have topological quantum number $1$ and $\psi$ respectively.   The results of such a calculation are shown in Fig.~\ref{fig:movingqhstogether}.    We see that the energy of moving two qh's together is always positive simply due to the Coulomb repulsion.  However, the energy is substantially greater when the two qh's are in the $|1\ra$ state compared to the $|\psi\ra$ state.  To our knowledge, this result was not predicted and may be attributed to the fact that the electron density vanishes in the $|1\ra$ state when two qh's approach each other, but remains non-zero in the $|\psi\ra$ state resulting in a more extended object \cite{Zikos}.   Thus our calculation makes the first mapping between a proposed measurement of the energy of two qh's and what this would indicate in terms of determining their topological quantum number.  (Obviously if one were moving a qp together with a qh, the $|1\ra$ state would have lower energy).   We also point out that this result may be significant in regards to the possibility of qh's condensing into daughter states as proposed in Refs.~\onlinecite{Bonderson} and \onlinecite{RandomHopping}.

The magnitude of the energy splitting of the two states, when two qh's
are very close to each other (within $\ell_0$), is measured to be roughly $0.01 e^2/\epsilon \ell_0$ which in a real system
corresponds to roughly 1 K, a rather small energy to be measured.  To
make matters worse, this measured energy should be considered to be an
upper bound, as mixing with states above the gap will be substantial
and could easily reduce this energy scale (the experimentally measured
gap itself is less than 1 K in the very best samples, although
theoretically without disorder the gap could be almost 2.5 K. See
Ref.~\onlinecite{RevModPhys} and therein).  Nonetheless, this
numerical work gives the first order of magnitude estimate for how
large the splitting due to topological quantum numbers is likely to be
compared to the overall Coulomb energy between the two qh's.

The decay length scales and energy scales that we calculate above are also extremely relevant to majorana tunneling.    Plugging in real numbers, we find that at a separation of about 0.1 micron, the energy diff erence will be about 80mK with the sign of the tunneling amplitude depending sensitively on the distance.  These scales have
implication to interferometry experiments\cite{Interferometry} where tunneling occurs
between edge and bulk as well as for majorana hopping problem where
tunneling occurs between many bulk qh's\cite{RandomHopping}.

To summarize, we have used Monte Carlo techniques to examine several
key properties of the MR wf with qh's.  Note that because our calculations do not incorporate  LL mixing
terms (which are expected to be small), they are equally applicable to the recently
proposed AntiPfaffian wf\cite{AntiPfaffian}.  We find that
both the unitary transformation associated with adiabatic transport
and the energy splitting of putatively degenerate states converge
exponentially with increasing distance between qp's, and we
explicitly extract the decay lengths.  Encouragingly, the
  decay lengths are on the order of a magnetic length which suggests
  that qp spacing should not be a barrier to physical implementation
  of topological operations.  Further we examine the energy splitting
that occurs when two qh's are moved together.  We find that the
$|1\rangle$ state of these two particles is of {\sl higher} energy and
we measure this energy splitting between $|1\ra$ and $|\psi\ra$.
Although this energy splitting is small, it gives experimentalists
another way to measure topological quantum numbers in these systems.
Many more details of this work will be presented in an upcoming
publication\cite{Zikos}.

We thank N.~Read, S.~M.~Girvin, P.~Bonderson, and J.~K.~Slingerland for helpful input.  
M.B.'s computer
time was supported by NSF grants DMR-0603369 and DMR-0653377.
G.Z. and N.E.B. are supported by US DOE Grant No. DE-FG02-97ER45639.


\begin{thebibliography}{14}
\bibitem{endnote1} A system with uncertain parity of $N$ would have degeneracy $2^{n_{qh}/2}$.  See Ref.~\onlinecite{RevModPhys}.
\end{thebibliography}
\end{document}